\documentclass[preprint,amsmath,amssymb,prb,floatfix,superscriptaddress]{revtex4}

\usepackage[dvips]{graphicx}
\usepackage{dcolumn}
\usepackage{bm}
\usepackage{mathrsfs}

\begin{document}
\title{Terahertz frequency magnetoelectric effect in Ni doped CaBaCo$_4$O$_7$}
\author{Shukai Yu}
 \affiliation{Department of Physics and Engineering Physics, Tulane University, 6400 Freret St., New Orleans, LA 70118, USA}
\author{C. Dhanasekhar}
 \affiliation{Cryogenic Engineering Centre, Indian Institute of Technology, Kharagpur 721302, India}
\author{Venimadhav Adyam}
 \affiliation{Cryogenic Engineering Centre, Indian Institute of Technology, Kharagpur 721302, India}
\author{Skylar Deckoff-Jones}
 \affiliation{Department of Physics and Engineering Physics, Tulane University, 6400 Freret St., New Orleans, LA 70118, USA}
\author{Michael K.L. Man}
 \affiliation{Femtosecond Spectroscopy Unit, Okinawa Institute of Science and Technology Graduate University, 1919-1 Tancha, Onna, Japan}
\author{Julien Madeo}
 \affiliation{Femtosecond Spectroscopy Unit, Okinawa Institute of Science and Technology Graduate University, 1919-1 Tancha, Onna, Japan}
\author{E Laine Wong}
 \affiliation{Femtosecond Spectroscopy Unit, Okinawa Institute of Science and Technology Graduate University, 1919-1 Tancha, Onna, Japan}
\author{Takaaki Harada}
 \affiliation{Femtosecond Spectroscopy Unit, Okinawa Institute of Science and Technology Graduate University, 1919-1 Tancha, Onna, Japan}
\author{Bala M.K. Mariserla}
 \affiliation{Femtosecond Spectroscopy Unit, Okinawa Institute of Science and Technology Graduate University, 1919-1 Tancha, Onna, Japan}
\author{Keshav M. Dani}
 \affiliation{Femtosecond Spectroscopy Unit, Okinawa Institute of Science and Technology Graduate University, 1919-1 Tancha, Onna, Japan}
\author{Diyar Talbayev}
 \email{dtalbayev@gmail.com}
 \affiliation{Department of Physics and Engineering Physics, Tulane University, 6400 Freret St., New Orleans, LA 70118, USA}

\date{\today}

\newcommand{\cbco}{CaBaCo$_4$O$_7$\:}
\newcommand{\cm}{\:\mathrm{cm}^{-1}}
\newcommand{\T}{\:\mathrm{T}}
\newcommand{\mc}{\:\mu\mathrm{m}}
\newcommand{\ve}{\varepsilon}
\newcommand{\dg}{^\mathtt{o}}

\begin{abstract}
We present a study of terahertz frequency magnetoelectric effect in ferrimagnetic pyroelectric \cbco and its Ni-doped variants.  The terahertz absorption spectrum of these materials consists of spin excitations and low-frequency infrared-active phonons.  We studied the magnetic-field-induced changes in the terahertz refractive index and absorption in magnetic fields up to 17 T.  We find that the magnetic field modulates the strength of infrared-active optical phonons near 1.2 and 1.6 THz.  We use the Lorentz model of the dielectric function to analyze the measured magnetic-field dependence of the refractive index and absorption.  We propose that most of the magnetoelectric effect is contributed by the optical phonons near 1.6 THz and higher-frequency resonances.  Our experimental results can be used to construct and validate more detailed theoretical descriptions of magnetoelectricity in CaBaCo$_{4-x}$Ni$_x$O$_7$.
\end{abstract}

\maketitle

\section{Introduction}
\cbco (CBCO) is a pyroelectric ferrimagnet with a frustrated magnetic structure that consists of alternating triangular and Kagome layers of magnetic Co ions\cite{caignaert:453,qu:917}.  It also belongs to a class of magnetoelectric multiferroics, which are materials that combine a magnetic order with electric polarization and exhibit strong coupling between the two.  In CBCO, the electric polarization is pyroelectric and is not switchable by application of an electric field\cite{johnson:045129}.  At the same time, strong magnetoelectricity has been reported in CBCO as it enters the ferromagnetic phase with $T_C\approx 60$ K: a large polarization change $\Delta P=17$ mC/m$^2$ happens due to the magnetic ordering\cite{caignaert:174403}.  The magnetically induced polarization change is much higher than the one reported in magnetic improper ferroelectrics, with GdMn$_2$O$_5$ showing the closest reported lower value of magnetically tuned polarization change\cite{lee:137203} of $\Delta P=5$ mC/m$^2$.  The record magnetoelectric coupling in CBCO is accompanied by a large modulation of the dielectric constant by magnetic field.  Near the magnetic phase transition, the static dielectric constant $\varepsilon$ is reduced by more than 15\% in magnetic field of 15 T\cite{singh:024410}.  Magnetoelectricity result from exchange striction\cite{singh:024410, caignaert:174403,johnson:045129}, and can potentially be useful in sensors/actuators or data storage storage applications. Thus, a detailed microscopic understanding of these effects is needed.

Terahertz (THz) spectroscopy has played a central role in unraveling the microscopic physics of magnetoelectricity.  THz frequency range (1 THz $\approx 4$ meV $\approx 33$ cm$^{-1}$) is usually home to spin excitations in antiferromagnets and low frequency optical phonons.  Detailed investigations of spin excitations in high magnetic field provide the experimental basis for the construction of accurate microscopic magnetic Hamiltonians\cite{talbayev:017202,talbayev:247601, standard:144422, nagel:257201}.  THz spectroscopy has also provided the observations of electromagnons\cite{pimenov:97,pimenov:014438,aguilar:060404,aguilar:047203,rogers:174407,takahashi:187201,chaix:157208,chaix:137201} which are electric-dipole-active spin excitations that receive their electric-dipole activity via microscopic magnetoelectric interactions.  In multiferroic manganites $R$MnO$_3$, THz-frequency electromagnons provided a detailed microscopic mechanism for exchange striction that governs the dynamic magnetoelectricity in these materials\cite{aguilar:047203}.  THz optical properties of undoped CBCO have been previously reported by Bord\'acs $et$ $al.$\cite{bordacs:214441}, who found sharp absorption lines in the 1-2 THz range due to spin and optical phonon excitations.  More excitingly, the authors found unidirectional THz absorption in CBCO, which results from dynamical magnetoelectric coupling\cite{kezsmarki:127203}.  In this article, we investigate the microscopic origin of the magnetoelectric effect in parent and Ni-doped CBCO.  We measure magnetic-field-induced
change in the THz absorption and refractive index.  We establish that the magnetic field modulates the oscillator strength of low-frequency infrared-active optical phonons, which results in a suppressed low-frequency refractive index and dielectric function.  This mechanism provides deeper insight in the microscopic exchange striction and can be used to construct accurate theoretical models of magnetoelectricity in CBCO. 

\begin{figure}[ht]
\begin{center}
\includegraphics[width=3in]{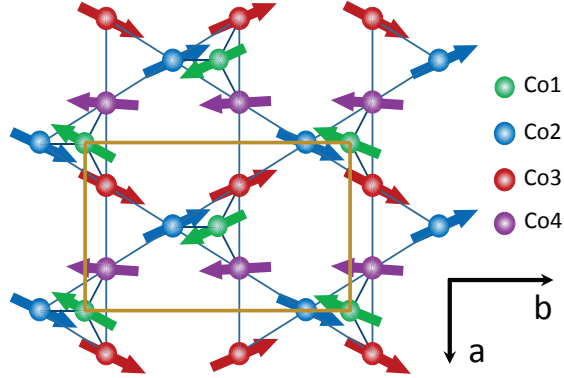}
\caption{\label{fig:magstruct}(Color online) Magnetic structure of \cbco.  Triangular layers are formed by the Co1 ions.  Kagome layers are formed by Co2, Co3, and Co4 ions.}
\end{center}
\end{figure}

\begin{figure}[ht]
\begin{center}
\includegraphics[width=4in]{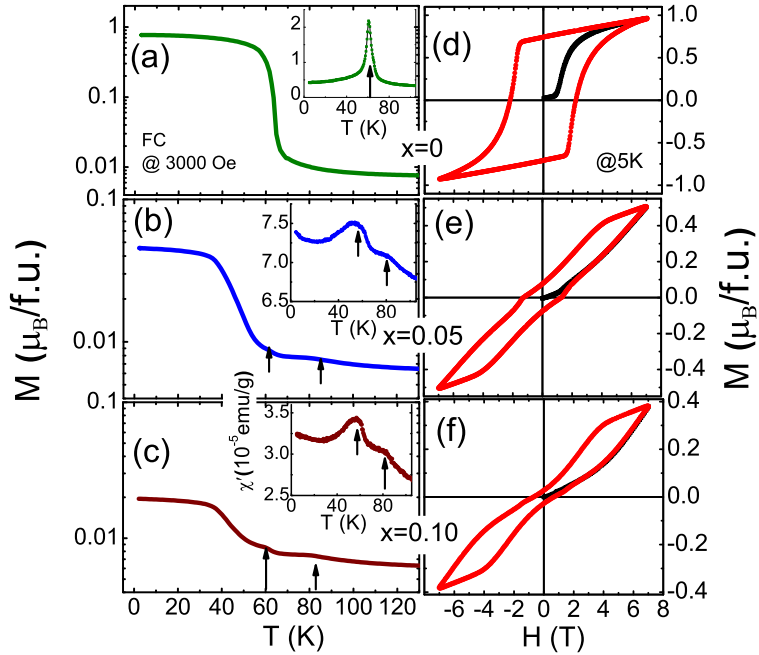}
\caption{\label{fig:magnetometry}(Color online) Temperature dependence of $dc$ magnetization (in logarithmic scale) (a-c) and magnetic hysteresis loops (d-f) in parent and Ni-doped CBCO. The insets in (a-c) show the $ac$ magnetization for the parent and Ni-doped CBCO with $H_{ac}=1$ Oe and $f=973$ Hz.}
\end{center}
\end{figure}

The magnetic structure of undoped CBCO is made of alternating Kagome and triangular layers of Co ions\cite{caignaert:453, caignaert:094417, singh:024410, caignaert:174403}.  The CBCO unit cell contains four inequivalent Co sites with equal ratio of Co valence states 2+ and 3+, which gives rise to ferrimagnetism.  The material's magnetic structure is reproduced in Fig.~\ref{fig:magstruct} following Caignaert $et$ $al.$\cite{caignaert:094417}. In the magnetic ground state, the magnetic moments of the four Co ions align almost collinearly and belong to the $ab$ plane, pointing mostly either along or opposite the $b$ axis.  The higher-valence ions and higher magnetic moments are found on Co1 and Co4 sites\cite{johnson:045129} (Fig.~\ref{fig:magstruct}).  Undoped CBCO samples used in this study show ferrimagnetic ordering at $T_C=60$ K.  Previous work on doped CBCO considered doping with Zn, which resulted in spectacular switching from ferrimagnetism to antiferromagnetism\cite{sarkar:232401,seikh:244106}.  Doping with Al causes a structural transition to hexagonal symmetry, as well as suppresses the magnetization\cite{zou:1}.  A dramatic suppression of ferrimagnetism was also observed upon Fe doping\cite{seikh:166}.  In Ni-doped CBCO (Fig.~\ref{fig:magstruct}), Co2 and Co3 ions are substituted with Ni ions to form the composition CaBaCo$_{4-x}$Ni$_x$O$_7$ with $x$=0.05 and 0.10.  The $dc$ and $ac$ magnetization measurements showed that both doped compositions exhibit magnetic phase transitions at 82 K in addition to the phase transition at 60 K already present in the parent compound (Fig.~\ref{fig:magnetometry}).  The $dc$ magnetization for the doped samples increases below 40 K and is consistent with $ac$ magnetization behavior. Both doped variants exhibit a saturation magnetization that diminishes from 0.95 $\mu_B$/f.u. in undoped CBCO to 0.50 $\mu_B$/f.u. ($x$=0.05) and 0.39 $\mu_B$/f.u. ($x$=0.10) in 7 T magnetic field (Fig.~\ref{fig:magnetometry}).  With increased doping, a butterfly shape of the $M(H)$ curves becomes more prominent (Fig.~\ref{fig:magnetometry}), which suggests increased antiferromagnetic interaction in doped CBCO\cite{dhanasekhar:01712}.  Low-temperature neutron powder diffraction confirmed that the Ni-doped CBCO retains the orthorhombic $Pbn2_1$ symmetry of the parent compound\cite{dhanasekhar:01712}.  Ni ions preferentially occupy Co2 and Co3 sites (Fig.~\ref{fig:magstruct}) as expected from their 2+ valency.  Neutron powder diffraction also allowed the determination of the magnetic ground state in which all magnetic moments point collinearly along the crystalline $a$ axis in the material with $x$=0.10 Ni content\cite{dhanasekhar:01712}.  Most interestingly, the switch to such collinear magnetic structure is accompanied by a switch to ferroelectric behavior, which partially motivates the present study\cite{dhanasekhar:01712}. 

\section{Experimental details}
Polycrystalline pellets of CaBaCo$_{4-x}$Ni$_x$O$_7$ (x=0, 0.05, and 0.10) were prepared by solid state reaction method\cite{dhanasekhar:76}.  X-ray powder diffraction  confirmed the phase purity of the samples at room temperature and their orthorhombic structure with $Pbn2_1$ space group. The pellets with Ni content of $x$=0, 0.05, and 0.10 had measured thicknesses of 0.491, 0.476, and 0.508 mm.

Zero-field THz time-domain spectroscopy (THz-TDS) was used to measure THz refractive index and absorption of the \cbco pellets in zero magnetic field.  The zero-field spectrometer is a home-built instrument that relies on optical rectification and electro-optic sampling for THz generation and detection\cite{silwal:092116}.  THz spectroscopy in high magnetic field (up to 17 T) was performed using a THz spectrometer based on photoconductive antennas for THz emission and detection.  The spectrometer was coupled to a superconducting solenoid magnet with optical access at the ends of the solenoid.  Polyethylene lenses were installed on the outer cold shield of the magnet to focus THz radiation onto the sample.  All THz measurements in magnetic field were performed in the Faraday geometry.

\section{Experimental results and discussion}

\begin{figure}[ht]
\begin{center}
\includegraphics[width=3in]{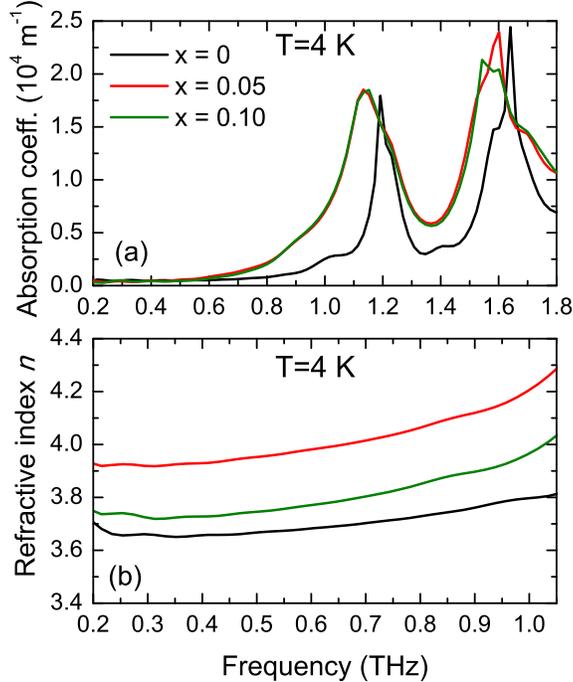}
\caption{\label{fig:thztds}(Color online) THz absorption (a) and refractive index (b) of \cbco pellets with different Ni content.}
\end{center}
\end{figure}

Figure~\ref{fig:thztds} shows the measured THz absorption and refractive index of \cbco at T=4 K.  The refractive index of undoped CBCO ($n\simeq3.7$) agrees with the value cited by Bord\'acs $et$ $al.$\cite{bordacs:214441}.  THz absorption of the undoped compound is also consistent with the previously reported results: we find two strong absorption lines at $1.2$ and $1.6$ THz (Fig.~\ref{fig:thztds}(a)).  The 1.2 THz resonance is attributed to $ab$-plane lattice vibrations, while the broader 1.6 THz resonance consists of two closely spaced $ab$-plane and $c$-axis phonon modes\cite{bordacs:214441}.  We also find weaker absorption resonances at 1.02 and 1.4 THz of spin wave origin\cite{bordacs:214441}.  Doping with Ni significantly broadens the 1.2 and 1.6 THz phonon absorption resonances (Fig.~\ref{fig:thztds}(a)).  Both resonances appear to be doublets consisting of a lower-frequency main absorption line and a higher-frequency shoulder.  We no longer observe the 1.4 THz spin wave resonance in the doped CBCO.  As this resonance is much weaker than the phonon absorption resonances, it may be "buried" under the broadened phonon resonances in the doped material.  The 1.02 THz spin wave resonance is shifted down to $\sim0.9$ THz in doped CBCO, as evidenced by the low-frequency shoulder in the absorption of the doped CBCO.  

\begin{figure}[ht]
\begin{center}
\includegraphics[width=3in]{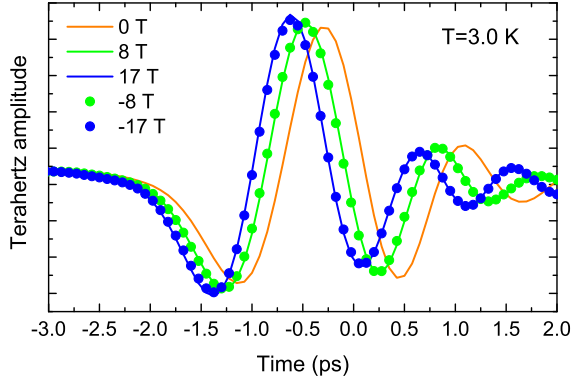}
\caption{\label{fig:thzpulse}(Color online) THz pulses transmitted by a \cbco pellet with $x$=0.05 Ni content at different applied magnetic fields.  THz pulses arrive earlier in higher magnetic fields, indicating a lower refractive index and dielectric constant.  The induced change in the refractive index is even in the direction of the applied magnetic field, i.e., does not change sign when magnetic field is reversed.}
\end{center}
\end{figure}

Figure~\ref{fig:thzpulse} shows the change in the transmitted time-domain THz pulse in CBCO with $x$=0.05 Ni content as magnetic field is applied along the light propagation direction (Faraday geometry).  The THz pulses arrive earlier for the higher applied field, which means that THz refractive index is lowered by the magnetic field.  Reversing the magnetic field direction also lowers the refractive index, indicating that the refractive index change is even in magnetic field.  When absorption is negligible, the refractive index is related to the real part of the dielectric function as $n(\omega)=\sqrt{\varepsilon_1(\omega)}$.  We use the notation $\tilde{n}(\omega)=n(\omega)+ik(\omega)$ and $\tilde{\varepsilon}(\omega)=\varepsilon_1(\omega)+i\varepsilon_2(\omega)$, so that $n(\omega)$ and $\varepsilon_1(\omega)$ are both real.  We interpret the field-induced change in refractive index as the modulation of the dielectric function $\varepsilon_1(\omega)$ by magnetic field, i.e., a THz frequency magnetoelectric effect in CBCO.  

\begin{figure}[ht]
\begin{center}
\includegraphics[width=3in]{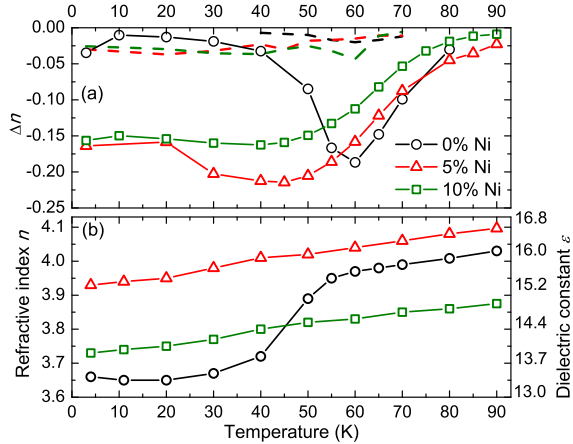}
\caption{\label{fig:deltan}(Color online) (a) Solid lines with symbols plot the magnetic field induced change in refractive index $\Delta n(\omega)=n(\omega,B)-n(\omega,0$\:T$)$ at frequency $\omega=0.4$\:THz and in applied field $B=17$\:T. Dashed lines plot the quantity $\Delta n_1(\omega)$, Eq. (\ref{eq:dn1}).  (b) Temperature dependence of refractive index $n(\omega)$ at frequency $\omega=0.4$\:THz and in zero magnetic field.  The vertical axis on the right hand side indicates the real part of the dielectric function at 0.4 THz calculated from $n(\omega)=\sqrt{\varepsilon_1(\omega)}$.}  
\end{center}
\end{figure}

To quantify the THz magnetoelectric effect, we measure the refractive index change $\Delta n(\omega)=n(\omega,B)-n(\omega,0$\:T$)$ at $\omega=0.4$\:THz and the highest applied magnetic field of $B=17$\:T (Fig.~\ref{fig:deltan}(a)).  The measured maximum relative change $\Delta n(\omega)/n(\omega)$ in undoped CBCO is $-5$\% at $T_c$ and can be translated into the relative change of the dielectric constant of $\Delta\varepsilon_1(\omega)/\varepsilon_1(\omega)=2\Delta n(\omega)/n(\omega)=-10$\%.  This magnitude of the THz magnetoelectric effect is comparable to the magnetoelectric effect in static dielectric permittivity $\varepsilon$ of $-16$\% in undoped CBCO pellets just below $T_c$ in 14 T magnetic field measured by Singh $et$ $al.$\cite{singh:024410}  These authors also find that the static dielectric permittivity $\varepsilon$ changes by $\Delta\varepsilon\simeq -2.7$ between 80 and 10 K, which matches our measured value $\Delta\varepsilon_1(\omega)=-2.7$ (Fig.~\ref{fig:deltan}(b)).  The temperature dependence of the static $\varepsilon$ exhibits a peak near $T_c$ in zero magnetic field\cite{singh:024410}.  We do not observe such peak in the temperature dependence of $\varepsilon_1(\omega)$ (Fig.~\ref{fig:deltan}(b)).  Dielectric measurements performed on large single crystals of CBCO indicate that the peak near $T_c$ is present only in the dielectric permittivity $\varepsilon$ along the $c$ axis\cite{caignaert:174403}.  No peak and only a sharp drop in permittivity is found along the $b$ axis.  The presence of the peak in the static permittivity $\varepsilon$ may be related to lower-frequency dynamics associated with critical fluctuations near $T_c$ or the dynamics of magnetic domains.  Despite the absence of the peak near $T_c$, the observed temperature and magnetic field behavior of the refractive index $n(\omega)$ and the corresponding $\varepsilon_1(\omega)$ at 0.4 THz in undoped CBCO result from the same fundamental physics that governs the static magnetoelectricity\cite{singh:024410,caignaert:174403}.

In doped CBCO with $x$=0.05 and 0.10, we find no strong signature of the magnetic phase transition in the temperature dependence of $n(\omega)$ and $\varepsilon_1(\omega)$ (Fig.~\ref{fig:deltan}(b)).  At a first glance, this would point to a diminished sensitivity of dielectric properties to magnetic structure.  However, we also find that the THz magnetoelectric effect is equally strong or even slightly enhanced in Ni doped CBCO, as evidenced in Fig.~\ref{fig:deltan}(a).  The onset of the magnetic field induced $\Delta n(\omega)$ occurs at higher temperature in doped CBCO, which then reaches is maximum at about 45 K and remains relatively unchanged below that temperature.  Thus, the magnetoelectric effect exists in a much wider temperature range in the doped CBCO compared to the undoped material.

\begin{figure}[ht]
\begin{center}
\includegraphics[width=3in]{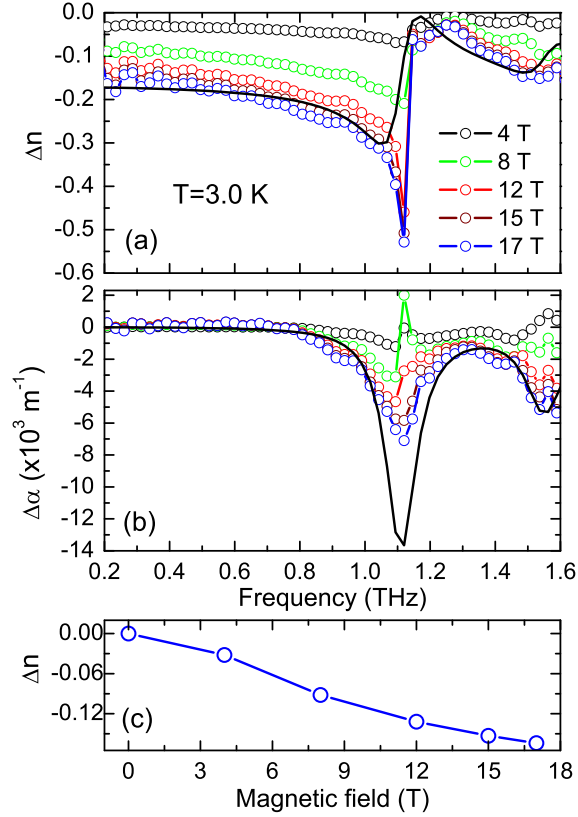}
\caption{\label{fig:dnvsb}(Color online) Change in the refractive index $\Delta n(\omega)$ (a) and absorption coefficient $\Delta\alpha(\omega)$ (b) in applied magnetic field in CBCO with $x$=0.05 Ni content. Solid black line shows the fit to the Lorentz model and Eqs. (\ref{eq:dnsim}) and (\ref{eq:dalphasim}).  (c) The magnetic field dependence of $\Delta n(\omega)$ at 0.4 THz.}
\end{center}
\end{figure}

We now consider the measured frequency dependences of $\Delta n(\omega)$ and $\Delta\alpha(\omega)$ (Fig.~\ref{fig:dnvsb}).  They were calculated from the measured frequency domain THz transmission spectra $S_B(\omega)$ and $S_0(\omega)$ collected with and without applied magnetic field, respectively:
\begin{eqnarray}
\label{eq:dn}
\Delta n(\omega)=\left( \arg S_B(\omega)-\arg S_0(\omega)\right) \frac{c}{\omega d},\\
\label{eq:dalpha}
\Delta\alpha(\omega)=\frac{-2\ln(S_B(\omega)/S_0(\omega))}{d},
\end{eqnarray}
where $c$ is the speed of light and $d$ is the pellet thickness.  Equation (\ref{eq:dalpha}) is valid when the relative change in index $n(\omega)$ is small.  Both quantities exhibit resonant behavior near the strong phonon absorption lines at 1.2 and 1.6 THz, (Fig.~\ref{fig:dnvsb}(a,b)), which we model by using the Lorentz oscillator model for the dielectric function
\begin{equation}
\label{eq:lorentz}
\tilde{\varepsilon}(\omega)=\varepsilon_\infty +\sum_{l=1,2}{\frac{f_l}{\omega_l^2-\omega^2-i\gamma_l\omega}},
\end{equation}
with $\omega_l$, $\gamma_l$, and $f_l$ and being the resonance frequency, relaxation rate, and oscillator strength of the $l$-th oscillator in the model.  Our model includes only two Lorentz oscillators, thus $l=1,2$.  $\varepsilon_\infty$ represents the higher-frequency dielectric response.  From (\ref{eq:lorentz}), we compute the field-induced change in refractive index and absorption under the simplifying assumption that the main effect of magnetic field is to modulate the oscillator strengths $f_1$ and $f_2$ of the phonon absorption lines at 1.2 and 1.6 THz.  A comparison of Figs.~\ref{fig:thztds}(a) and~\ref{fig:dnvsb}(b) shows that such assumption is justified.  We also assume that the field induced relative change in $n(\omega)$ is small.  We calculate $\Delta n(\omega)$ and $\Delta\alpha(\omega)$ as 
\begin{eqnarray}
\label{eq:dnsim}
\Delta n(\omega)=\Delta n_\infty + \frac{1}{2 n(\omega)}\sum_{l=1,2}{\frac{(\omega_l^2-\omega^2)\:\Delta f_l}{(\omega_l^2-\omega^2)^2 + \gamma_l^2\omega^2}},\\
\label{eq:dalphasim}
\Delta\alpha(\omega)=\frac{\omega}{c n(\omega)}\sum_{l=1,2}{\frac{\gamma_l\omega\:\Delta f_l}{(\omega_l^2-\omega^2)^2 + \gamma_l^2\omega^2}},
\end{eqnarray}
where $\Delta f_l$ is the field-induced change in oscillator strengths and $\Delta n_\infty$ is a constant describing the field-induced change in the higher-frequency refractive index and dielectric function.

We now use equations (\ref{eq:dnsim}) and (\ref{eq:dalphasim}) to fit the experimental frequency dependence of the field-induced changes in refractive index and absorption with $\Delta n_\infty$, $\Delta f_l$, $\omega_l$, and $\gamma_l$ ($l=1,2$) as fitting parameters.  The measured $\Delta n(\omega)$ and $\Delta\alpha(\omega)$ exhibit large uncertainties near the two resonance frequencies $\omega_1/2\pi\approx 1.1$ THz and $\omega_2/2\pi\approx 1.6$ THz, as well as at frequencies above 1.6 THz, due to strong sample absorption and very small amount of measured light (Figs.~\ref{fig:dnvsb}(a,b)).  (Experiments with thinner samples may potentially resolve the high-frequency parts of the spectrum.) Thus, we restrict the least-squares fitting procedure to frequencies below 1.5 THz; we also exclude the data points in the immediate vicinity of the $\omega_1$ resonance.  We fit both $\Delta n(\omega)$ and $\Delta\alpha(\omega)$ simultaneously.  An example of the resulting best fit is shown in Figs.~\ref{fig:dnvsb}(a,b) for the data on Ni-doped CBCO ($x$=0.05) at $T=3$ K and $B=17$ T.  We note the disagreement between measurement and calculation in the vicinity of the Lorentz oscillator $\omega_1$.  The calculation overestimates the measured $\Delta\alpha(\omega)$ and underestimates the measured $\Delta n(\omega)$.  Nonetheless, the "aggregate" change in the refractive index and the absorption levels  above and below the $\omega_1$ resonance are well captured by the model.  One possibility for the discrepancy near $\omega_1$ is the uncertainty in measurement data; another possibility is a non-Lorentzian shape of the phonon mode due to the polycrystalline nature of the sample.  Effective medium theory may potentially provide a better description of the line shape.

The frequency dependence of the THz magnetoelectric effect (Fig.~\ref{fig:dnvsb}) provides a new and unique insight in the physics of magnetoelectric coupling in CBCO and its Ni doped variants.  In the Lorentz model description of the dielectric response, the static permittivity $\varepsilon$ in zero field is the sum of distinct contributions of individual Lorentz oscillators dependent on their oscillator strengths, (Eq. (\ref{eq:lorentz})).  (The higher-frequency dielectric constant $\varepsilon_\infty$ in Eq. (\ref{eq:lorentz}) must also be interpreted as a sum of contributions of higher-frequency electric dipole transitions.)  Similarly, the magnetic field induced change in $\varepsilon$ can be described as a result of the modulation of individual Lorentz oscillator responses by magnetic field.  The Lorentz oscillators in our model are the phonon modes at 1.2 and 1.6 THz, while the contributions of higher-frequency electric dipole transitions are accounted for by the constants $\varepsilon_\infty$ and $\Delta n_\infty$.  First-principles calculations indicate that exchange striction can account for the observed magnetoelectric effect in CBCO\cite{johnson:045129}.  Exchange striction refers to the modulation of the magnetic exchange interaction strength (e.g., the Heisenberg exchange constants $J$) with the corresponding bond length between the interacting magnetic ions.  Through exchange striction, spontaneous magnetic ordering and/or the application of magnetic field can modulate bond lengths, unit cell parameters, and phonon mode characteristics. The modulation of the phonon mode strengths by magnetic field leads directly to the THz and static magnetoelectric response, as described by Eqs. (\ref{eq:lorentz})-(\ref{eq:dalphasim}).  Thus, our experimental results provide evidence on which phonon modes contribute the most (or the least) to the magnetoelectric effect via the exchange striction.

Our data and fitting allow us to separate the contribution of the lowest observed phonon mode near 1.2 THz from the combined contributions of the higher frequency phonons.  While we also clearly observe the 1.6 THz phonon modes in our zero-field spectra (Fig.~\ref{fig:thztds}), this mode occurs right at the edge of the measurement frequency window in the magnetic-field dependent spectra (Fig.~\ref{fig:dnvsb}).  The fitting of the magnetic field spectra was performed using Eqs. (\ref{eq:dnsim}) and (\ref{eq:dalphasim}) that included both the 1.2 and 1.6 THz phonon modes.  From these fits, we determined the quantities 
\begin{eqnarray}
\label{eq:dn1}
\Delta n_1(\omega)= \frac{\Delta f_1}{2n(\omega)\omega_1^2},\\
\label{eq:dn2}
\Delta n_2(\omega)=\Delta n_\infty + \frac{\Delta f_2}{2n(\omega)\omega_2^2}.
\end{eqnarray} 
Their sum is equal to the total field induced change in refractive index at a low frequency, such as $\Delta n(\omega)$ at 0.4 THz shown in Fig.~\ref{fig:deltan}: $\Delta n_1(\omega)+\Delta n_2(\omega) = \Delta n(\omega)$.  The relative contribution of $\Delta n_1(\omega)$ to $\Delta n(\omega)$ is much smaller than the contribution of $\Delta n_2(\omega)$.  This is illustrated in Fig.~\ref{fig:deltan}(a), where $\Delta n_1(\omega)$ is plotted using dashed lines.  Thus, we propose that the magnetoelectric effect in CBCO is determined mostly by the high-frequency phonon modes at 1.6 THz and above.  Our data do not allow us to reliably separate the contributions of the two terms on the right hand side of Eq. (\ref{eq:dn2}), as the 1.6 THz phonon occurs right at the edge of the frequency window.  Our proposal is consistent with the fact that the strongest static magnetoelectric effect in CBCO happens along the $c$ axis\cite{caignaert:174403}, since the 1.2 THz phonon is attributed to the $ab$-plane lattice vibrations\cite{bordacs:214441}.  To gain further insight in the nature of magnetoelectricity in CBCO, a theoretical calculation and assignment of the observed phonon frequencies and modes is needed. 

\section{Conclusions}
We have investigated the THz-frequency magnetoelectric effect in parent and Ni-doped CBCO.  In the parent compound, the temperature and magnetic field dependence of the dielectric function $\varepsilon_1(\omega)$ at 0.4 THz closely resembles the behavior of the static dielectric constant\cite{singh:024410} (Fig.~\ref{fig:deltan} and Fig.~\ref{fig:dnvsb}(c)).  Thus, the THz and static magnetoelectric effects share the same microscopic origin.  The magnetoelectric effect in undoped CBCO is strongest near the ferrimagnetic phase transition temperature $T_C$ and becomes negligible toward zero temperature.  Doped CBCO ($x$=0.05 and 0.10) exhibits an additional magnetic transition at 82 K together with the 60 K transition.  The exact nature of these transitions remains to be clarified.  In doped CBCO, the THz magnetoelectric effect appears near the 82-K phase transition and steadily grows as the temperature is lowered toward 40 K, where it reaches maximum strength that remains practically unchanged at lower temperatures (Fig.~\ref{fig:deltan}).  The magnetic-field behavior of frequency-dependent THz absorption and refractive index (Fig.~\ref{fig:dnvsb}) points to a very specific mechanism for the magnetoelectric effect: a modulation of the strength of infrared-active phonon modes by magnetic field.  Our Lorentz model analysis allows us to conclude that the lowest observed optical phonon near 1.1-1.2 THz, an $ab$ plane lattice vibration\cite{bordacs:214441}, provides only a small contribution to the magnetoelectric effect.  We propose that the phonon modes near 1.6 THz and higher-frequency infrared-active resonances together provide the main contribution to the observed THz magnetoeletric effect.  Our present measurement does offer any detail about the nature of the possible higher-frequency modes, which are described by the $\Delta n_\infty$ term in Eq. (\ref{eq:dn2}); e.g., spectral weight modulation of electronic transitions by magnetic or orbital ordering may also contribute\cite{kovaleva:147204} to $\Delta n_\infty$.  The microscopic origin of magnetoelectricity  in CBCO is rooted in exchange striction\cite{singh:024410,johnson:045129}.  Our experimental results offer a firm basis for more refined future theoretical descriptions of exchange striction and magnetoelectricity in CBCO that account for the observed low-frequency phonon spectrum and its modulation by magnetic field.

\section*{Acknowledgments}
The work at Tulane University was supported by the NSF Award No. DMR-1554866.  The authors from IIT Kharagpur acknowledge DST, India for FIST project and IIT Kharagpur funded VSM SQUID magnetometer.


\end{document}